\documentclass[showpacs, twocolumn]{revtex4-1}
\usepackage{graphicx}
\usepackage{amsmath}
\usepackage{appendix}
\usepackage{array}
\usepackage{multirow}

\begin{document}

\title{ Weakly interacting Bose gases with generalized uncertainty principle: Effects of quantum gravity}

\author{Abdel\^{a}ali Boudjem\^{a}a}

\affiliation{Department of Physics, Faculty of Exact Sciences and Informatics, Hassiba Benbouali University of Chlef, P.O. Box 78, 02000, Ouled-Fares, Chlef, Algeria.}
\email {a.boudjemaa@univ-chlef.dz}


\begin{abstract}

We investigate quantum gravity corrections due to the generalized uncertainty principle on three-dimensional weakly interacting Bose gases at both zero and finite temperatures 
using the time-dependent Hatree-Fock-Bogoliubov theory. 
We derive useful formulas for the depletion, the anomalous density and some thermodynamic quantities such as the chemical potential, 
the ground-state energy, the free energy, and the superfluid density.
It is found that the presence of a minimal length  leads to modify the fluctuations of the condensate and its thermodynamic properties 
in the weak and strong quantum gravitational regimes.
Unexpectedly, the interplay of quantum gravity effects and quantum fluctuations stemming from interactions may lift both the condensate and the superfluid fractions.
We show that quantum gravity minimizes the interaction force between bosons leading to the formation of ultradilute Bose condensates. 
Our results which can be readily probed in current experiments may offer a new attractive possibility to understand gravity in the framework of quantum mechanics.

\end{abstract}

\maketitle

\section{Introduction}

Two crowing intellectual achievements of modern physics that exquisitely explain how nature works: quantum mechanics and theory of general relativity. 
Attempts to reconcile these incompatible theories usually involve quantizing gravity to formulate a theory of quantum gravity (QG) (see e.g. \cite{Penr}).
However, no experiment today provides evidence that gravity has a quantum mechanical origin.
The most serious obstacle is the incredible energies needed $\sim 10^{28}$ev (Planck energy) or equivalently a length-scale of the order of
the Planck length $\approx 10^{-35}$m.
Given current technologies, a direct observation of such intriguing QG effects would likely require to build a particle accelerator bigger than our galaxy.

With recent progress in quantum information science, the topic of experimentally testing QG has been gaining renewed interest.
Several proposals based on quantum information science such as quantum entanglement between two microspheres \cite{Kaf,Bose, Mar,Kris,Mar1} 
and non-Gaussianity in matter \cite{Howl} have been used to witness QG.
Ultracold atoms including macroscopic Bose-Einstein condensates  (BECs) offer another promising possibility of testing the quantum nature of the gravitational field
due to their extraordinary degree of control and sensitivity to ultraweak forces \cite{Howl, Shir, Bris, Bris1, Das2, Hans,Jaf,Simon}.
An additional advantage in using ultracold gases is that the electromagnetic interactions are adjustable by means of an external magnetic or optical fields \cite{Howl, Hans}.

Numerous approaches to QG such as string theory and loop QG, as well as black hole physics predict a minimum measurable length in nature, 
below which no other length can be observed. 
One of the most intriguing aspect linked to the existence of such a minimum length is the modification of the Heisenberg uncertainty principle
to a generalized uncertainty principle (GUP) \cite{Mag,Kempf}.  This latter predicts corrections to diverse quantum phenomena 
\cite{Mag,Kempf,Scar,Chang,Ali,Spre, Piko, Hus,Pedr,Feng,Shab, Gec, Scar1,Bra,Das3,Casa}.
Furthermore, based on minimum observable length, implications of QG on the statistical properties of ideal Bose gases have been widely investigated 
(see, e.g., \cite{Fit, VaK,Zhang, Cast, Li, Sanj, Das} and references therein).

However, to the best of our knowledge, the effects of QG on weakly interacting Bose gases remain rarely examined.
The model of weakly interacting Bose gases which is universal is extremely interesting since
it simultaneously covers the low-energy edge of the effective theory, and the high-energy  physics \cite{Volov}.
One remarkable property of such quantum ensembles is their dilutness which enables us to treat problems either completely, or perturbatively or even numerically in reasonable times, 
opening a feasible route for testing whether or not the gravitational field displays quantum properties. 

The aim of this paper is to investigate the effects of QG due to the GUP on weakly interacting homogeneous Bose gases at both zero and finite temperatures. 
To achieve this goal, we use  the time-dependent Hatree-Fock-Bogoliubov (TDHFB) theory \cite {Boudj, Boudj1,Boudj2}. 
Basically, the TDHFB theory is a self-consistent approach describing the dynamics of ultracold Bose gases at any temperatures which involves interactions 
between the condensate and the thermal cloud.
The TDHFB equations offer an elegant starting point to treat many-body dynamics and have been successfully applied to a wide variety of problems 
\cite {Boudj, Boudj1,Boudj2,Boudj3,Boudj4,Boudj5,Boudj6,Boudj7, Boudj8,Boudj9,Boudj10,Boudj11, Boudj12,Boudj13,Boudj14,Boudj16}.  
They could also provide an ideal alternative dynamical model involving the cosmological constant for understanding 
the accelerated expansion of the universe caused by dark matter.

When considering the effects of minimal length, the density of states is modified giving rise to substantially affect the Bogoliubov dispersion relation,
the quantum fluctuations, and the thermodynamic properties of the system (see e.g.\cite{VaK,Zhang,Das}).
In this work we compute correction terms arising from a natural deformation of QG governed by the HFB equations based on a minimal length framework to the 
condensed depletion, the anomalous density, the chemical potential, the ground-state energy, the free energy, and the superfluid density.
One key hurdle routinely encountered while implementing anomalous correlations and ground-state energy is the ultraviolet divergences caused by the short-range contact potential. 
In this regard, a number of techniques have been used to cure these difficulties such as the renormalization of the coupling constant \cite {Beleav,Griffin, peth,Boudj15} 
and the dimensional regularization \cite{Boudj, Anders,Yuk, Boudj17}.
Here, we show that the presence of a minimal length scale in the HFB formalism regularizes naturally the coupling constant (i.e. it damps short distance modes), 
and hence overcomes the well-known ultraviolet divergence problem.

At zero temperature, we find that QG effects may reduce both the quantum fluctuations and the thermodynamic quantities.
At finite temperature, corrections due to weak QG effects to the condensate thermal fluctuations, the free energy and the normal density of the superfluid are obtained analytically. 
We point out that these amendments are significant only at temperatures well below the transition. 
On the other hand, effects of strong QG are calculated  using  numerical simulations.
The results reveal that QG which behaves itself as quantum correlations, provides an additional attractive term which
competes with repulsive mean-field and Lee-Huang-Yang (LHY) energies (resulting from quantum and thermal fluctuations)  leading to lower the above observables of the condensate.
Crucially, we demonstrate that the presence of gravitational effects may strikingly enhance the condensed and the superfluid densities.
The validity criterion of the present HFB theory is accurately established at both zero and finite temperatures.
Furthermore, by measuring the corrections in the quantum depletion, we can constrain the deformation parameter and the minimum measurable length.

The rest of the paper is organized as follows. In section \ref {model}, we introduce the fundamental concepts of  the TDHFB model 
with GUP implying the existence of a minimal length. We derive in addition the essential formulas to tackle the problem under investigation.
The effects of QG on the quantum and thermal fluctuations, the thermodynamics, and on the superfluidity in BEC are deeply discussed in sections \ref{flcu}  and \ref{Therm}, respectively.
In Sec.\ref{Exp} we discuss the experimental relevance of our predictions.
Our conclusions are drawn in section \ref{conc}.

\section{TDHFB Theory with GUP}  \label{model}

We consider a three-dimensional dilute Bose gas with the atomic mass $m$ at a temperature $T$.
We make the plausible assumptions that bosons are weakly interaction via a contact potential  $V ({\bf x}-{\bf x'})=g  \delta ({\bf x}-{\bf x'})$,
where $g=(4\pi \hbar^2/m) a$ is the coupling constant  with $a$ being the $s$-wave scattering length. 
The Hamiltonian of such a system reads 
\begin{align} \label{eq4}
\hat H &=\int d{\bf x} \, \hat\psi^\dagger ({\bf x}) \left[h^{sp} +\frac{g}{2} \hat \psi^\dagger ({\bf x})\hat \psi ({\bf x})\right]\hat\psi({\bf x}), 
\end{align}
where  $\hat\psi^\dagger$ and  $\hat\psi$ are the boson destruction and creation field operators, respectively, satisfying the usual canonical commutation rules 
$[\hat\psi({\bf x}), \hat\psi^\dagger (\bf x')]=\delta ({\bf x}-{\bf x'})$.
The single particle Hamiltonian is defined by $h^{sp}=-(\displaystyle\hbar^2/\displaystyle 2m) \Delta + V({\bf x})-\mu$, where $V({\bf x})$ is the external potential
and $\mu$ is the chemical potential.

At finite temperature, we usually perform our analysis in the mean-field framework relying on the TDHFB equations.
These latter are based on the time-dependent Balian-V\'er\'eroni variational principle \cite{BV} with Gaussian trial time-dependent density operator $D(t)$. 
This ansatz is associated with the partition function ${\cal Z}$, the one-boson field expectation values 
$\langle \hat{\psi}\rangle ({\bf x},t)$, $\langle \hat{\psi}^\dagger\rangle ({\bf x},t)$, and the single-particle density matrix $\rho ({\bf x,x'},t)$ \cite{Boudj1}.
Upon inserting these variational parameters into the BV action, one obtains the TDHFB equations \cite {Cic,Ben, Boudj}
\begin{equation} \label {TDH1}
i\hbar \frac{d  \Phi}{d t} =\frac{d{\cal E}}{d \Phi},
\end{equation}
\begin{equation} \label {TDH2}
i\hbar \frac{d \rho}{d t} =-2\left[\rho, \frac{d{\cal E}}{d\rho} \right],
\end{equation}
where ${\cal E}=\langle \hat H\rangle$ is the energy of the system, and the single particle density matrix of a thermal component  is defined as:
$$
\rho=\begin{pmatrix} 
\langle \hat{\bar{\psi}}^\dagger\hat{\bar{\psi}}\rangle & -\langle\hat{\bar{\psi}}\hat{\bar{\psi}}\rangle\\
\langle\hat{\bar{\psi}}^\dagger\hat{\bar{\psi}}^\dagger\rangle& -\langle\hat{\bar{\psi}}\hat{\bar{\psi}}^\dagger\rangle
\end{pmatrix},
$$
where $\hat{\bar \psi}({\bf x})=\hat\psi({\bf x})- \Phi({\bf x})$ is the noncondensed part of the field operator with $\Phi({\bf x})=\langle\hat\psi({\bf x})\rangle$ 
being the condensate wave-function.
Equations  (\ref{TDH1}) and (\ref{TDH2}) imply that the energy ${\cal E}$ is conserved when the Hamiltonian $H$ does not depend explicitly on time.
They constitute a closed set of equations for a condensate coexisting with a thermal cloud and a pair anomalous density. 

An important feature of the TDHFB  formalism is that it allows unitary evolution of  $\rho$.
Then the conservation of the Von Neumann entropy $S=Tr D\ln D$ yields
\begin{align} \label{Invar}
( I-1)/4=\rho (\rho+1),
\end{align} 
where $I$ is often known as the Heisenberg invariant \cite{Cic,Ben, Boudj}. It represents the variance of the number of noncondensed particles. 
For pure state and at zero temperature, one has $I=1$.

Many QG models predict a minimal uncertainty length.
The simplest form of GUP relation which implies the appearance of a nonzero minimal uncertainty is proposed as \cite{Kempf}:
\begin{equation}\label {GUP}
\Delta X \Delta P \geq  \frac{\hbar}{2}  \left[ 1+ \beta (\Delta P)^2\right],
\end{equation}
where  $\beta >0$ is the deformation parameter which is often expressed in terms of the dimensionless parameter  $\beta_0$,
 and  the Planck length $l_p$ as: $\beta= \beta_0 l_p^2/\hbar^2=\beta_0/(M_p c)^2$, where $M_p=\sqrt{\hbar c/G}$ is the Planck mass with
$G$ being the gravitational constant and $c$ denotes the speed of light in vacuum. 
Current experiments can set upper bounds on the GUP parameter. For instance,  
the standard model of high-energy physics implies that $\beta_0 < 10^{34}$ \cite{Das2}. 
According to the same reference \cite{Das2}, the scanning tunneling microscope delivers the best one $\beta_0 < 10^{21}$ \cite{Das2}. 
Other upper bounds have been provided by different approaches, namely the Lamb shift and Landau levels \cite{Das2}, optical systems \cite{Bra},  
the light deflection and perihelion precession \cite{Scard}, cold atoms \cite{Gao}, and gravitational systems \cite{Feng1,Nev,DD}.
However, in the present work we will address QG effects for arbitrary $\beta_0$.
Evidently, for $\beta=0$, the standard Heisenberg uncertainty principle is recovered. 

Equation (\ref{GUP}) immediately leads to the definition of the minimum measurable length
\begin{equation}\label {GUP0}
\Delta X_{\text{min}}= \hbar \sqrt{3 \beta}= \sqrt{3 \beta_0} l_p.
\end{equation}
In the case of mirror-symmetric states (i.e. $\langle \hat P\rangle =0$), it is possible to obtain Eq.(\ref{GUP}) from the modified commutation relation:
\begin{equation}\label {GUP1}
[\hat X, \hat P]= i\hbar \left( 1+ \beta  P^2\right).
\end{equation}
According to Eqs.(\ref{GUP}) and (\ref{GUP1}), the deformed density of states can be given by \cite{Kempf,Chang}
\begin{equation}\label {GUP2}
{\cal D} (P) dP= \frac{V} {2 \pi^2 \hbar^3} \frac{P^2 dP} {(1+\beta P^2)^3}.
\end{equation} 
Equation (\ref{GUP2}) shows that the GUP may modify the statistics and the thermodynamics of a weakly interacting Bose gas. 

From now on, we focus on uniform Bose gases and assume that the dynamics of the thermal cloud and the anomalous density is not important at low temperatures.
Effects of QG in quantum systems are implied by the GUP given in Eq.(\ref{GUP}).
Therefore, the above TDHFB equations take the explicit form :
\label {model}
\begin{equation}\label {T:DH}
i\hbar \frac{d  \Phi}{d t} = \left[ E_P+g (n_c+2\tilde n+\tilde m)-\mu\right]\Phi,
\end{equation}
where $E_P=P^2/2m$ is the energy of free particle, $n_c=|\Phi|^2$ is the condensed density, $\tilde n=\langle \hat{\bar{\psi}}^\dagger\hat{\bar{\psi}}\rangle$ 
is the noncondensed density, and $\tilde m=\langle\hat{\bar{\psi}}\hat{\bar{\psi}}\rangle$ accounts for the anomalous correlation. The total density is given by $n=n_{c}+\tilde n$. 
Equation (\ref{T:DH}) is the generalized Gross-Pitaevskii equation which allows us to study the static and the dynamics of the condensate 
in terms of the paradigms developed in the GUP framework.

The chemical potential reads
\begin{equation}\label {chim}
\mu = g (n_c+2\tilde n+\tilde m)=gn +g(\tilde n+\tilde m).
\end{equation}
The term $g(\tilde n+\tilde m)$ represents the quantum corrections to the chemical potential owing to the BEC fluctuations.

The Bogoliubov excitations energy can be obtained by linearizing Eq.(\ref{T:DH}) employing the standard transformation:
$\Phi ({\bf P},t)=\sqrt{n_c}+ u_P  e^{-i\varepsilon_P t}+v_P e^{i\varepsilon_P t}$, where $\sqrt{n_c}=\Phi_0=\Phi_0^*$ is the equilibrium solution,
and $u_P,v_P =(\sqrt{\varepsilon_P /E_P} \pm \sqrt{E_P/\varepsilon_P})/2$ are the  Bogoliubov quasiparticle amplitudes \cite{Bog}.
This gives
\begin{equation}\label{BogR} 
\varepsilon_P= \sqrt{ (P^2/2m)^2 + c_s^2 P^2},
\end{equation}
where $c_s= \sqrt{g (n_c+\tilde m)/m}$ is the sound velocity.
For small momenta $P\rightarrow 0$, the Bogoliubov dispersion relation is phonon-like $\varepsilon_P= c_s P$ (quanta of sound waves). 
For the curl-free superfluids such sound waves play the role of "gravitational waves" \cite{Volov}. 
In the opposite limit $P\rightarrow \infty$, the excitations spectrum (\ref{BogR}) reduces to the free particle law : $\varepsilon_P=E_P$.

The effects of the minimal length on the behavior of quantum and thermal fluctuations of BEC can be calculated from
Eq.(\ref{TDH2}) which can be written in momentum space as:
\begin{equation}\label{T:DH2} 
\tilde n=\frac{1}{4 \pi^2 \hbar^3}\int \frac{P^2d  P} {(1+\beta P^2)^3} \left[\frac{E_P+ mc_s^2} {\varepsilon_P} \sqrt{I_P}-1\right],
\end{equation}
and
\begin{equation}\label{T:DH3} 
\tilde m=-\frac{1}{4 \pi^2 \hbar^3}\int \frac{P^2d  P} {(1+\beta P^2)^3} \frac{mc_s^2 } {\varepsilon_P} \sqrt{I_P}.
\end{equation}
In the absence of the GUP ($\beta=0$), the anomalous density (\ref{T:DH3}) becomes ultraviolet divergent due to the use of the short-range contact potential.
To circumvent this problem, we renormalize the coupling constant $g$ and introduce the Beliaev-type second-order corrections
$g_R ({\bf P})=g + g^2\int_{P<P_c} \frac {d{\bf P} }{(2\pi \hbar)^3} \frac{1}{2E_P}$ \cite {Beleav,peth,Boudj15}. Another strategy to cure such an ultraviolet divergence is
the dimensional regularization that is asymptotically accurate for weak interactions  \cite{Boudj, Anders,Yuk, Boudj17}.

The Heisenberg equality (\ref{Invar}) serves to provide a useful relation between the noncondensed (\ref{T:DH2}) and anomalous (\ref{T:DH3}) densities \cite{Cic, Boudj}:
\begin{equation}\label {heis}
I_P= (2\tilde{n}_P+1)^2-|2\tilde{m}_P|^2= \coth^2\left(\varepsilon_P/2T\right),
\end{equation}
here we used the identity $2N_P(x)+1= \coth (x/2)$ with $N_P= [\exp{(\varepsilon_P/T)}-1]^{-1}$ being occupation numbers for the excitations.
Equation (\ref{heis}) enables us to determine in a very useful way the critical temperature and the superfluid density of Bose quantum liquids (see below).

\section {Fluctuations} \label{flcu}

Our target now is to calculate corrections owing to the effects of QG governed by the HFB to the condensed depletion and the anomalous density at both zero and finite temperatures. 
Let us assume the limit $\tilde{m}/n_c \ll 1$,  which is valid at low temperature 
and necessary to ensure the diluteness of the system \cite{Boudj,Yuk}. Therefore, the sound velocity reduces to $c_s= \sqrt{g n_c/m}$.

\subsection{Quantum fluctuations}

\begin{figure}
\centering 
\includegraphics[scale=0.8, angle=0] {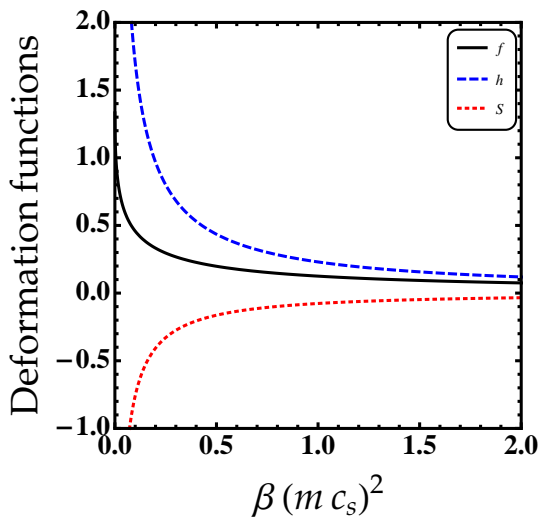}
\caption { Deformation functions $f(\beta)$, $h(\beta)$ and $S(\beta)$ which govern
the dependence of the condensate depletion, the anomalous fraction and the ground-state energy corrections  vs. the deformation parameter $\beta$ in units of $(mc_s)^2$.}
\label{Fluc}
\end{figure}

The noncondensed density can be straightforwardly evaluated from integral (\ref{T:DH2}). One finds 
\begin{equation}\label{T:DH22} 
\tilde n=\frac{(mc_s)^3}{4 \pi^2 \hbar^3}  f(\beta),
\end{equation}
where the deformation function $f(\beta)$ is given by
\begin{widetext}
\begin{align}\label{DF1} 
f(\beta) &= \frac{1} {16 (1-4 m^2c_s^2\beta )^{5/2} (m^2c_s^2\beta)^{3/2}} 
\bigg \{ \sqrt{1-4 m^2c_s^2 \beta } \bigg[4 \sqrt{m^2c_s^2\beta } \big(2 m^2c_s^2\beta\, (8 m^2c_s^2\beta -1)+1\big)-\pi  (1-4 m^2c_s^2\beta )^2\bigg] \\
&+ (2-20 m^2c_s^2\beta ) \arccos \left(2 \sqrt{m^2c_s^2 \beta}\right) \bigg \}, \nonumber
\end{align}
\end{widetext}
which is unrelated to atom mass $m$.
Importantly, for $\beta \rightarrow 0$, $f(\beta) \approx 4/3$ (see Fig.(\ref{Fluc}) solid line), we reproduce the result  of a dilute Bose gas without QG,    
$\tilde n=(mc_s)^3/(3 \pi^2 \hbar^3)$. 
In terms of a small parameter of the theory the depletion takes the form $\tilde n/n_c=8 \sqrt{n_c a^3/\pi}/3$, where the condensed density $n_c$ which constitutes our corrections 
with respect to the Bogoliubov results \cite{Bog}, appears as a key parameter instead of the total density $n$.

Thanks to GUP the anomalous density (\ref{T:DH3}) does not suffer from ultraviolet divergences. 
Consequently, integral (\ref{T:DH3}) yields 
\begin{equation}\label{T:DH33} 
\tilde m=-\frac{(mc_s)^3}{4 \pi^2 \hbar^3}  h(\beta),
\end{equation}
where  the QG correction function $h(\beta)$ is given by
\begin{align}\label{DF2} 
h(\beta) &=- \frac{1} {4 (1-4 m^2c_s^2\beta )^{5/2} \sqrt{ m^2c_s^2\beta}} \bigg[(10-16  m^2c_s^2 \beta ) \\
&\times \sqrt{ m^2c_s^2 \beta (1-4  m^2c_s^2 \beta) }-3 \arccos\left(2 \sqrt{ m^2c_s^2 \beta }\right) \bigg]. \nonumber
\end{align}
Again for $\beta \rightarrow 0$, $h(\beta) \approx 3\pi/(8 \sqrt{m^2c_s^2\beta})-4$, thus, the anomalous density  (\ref{T:DH33}) reduces to
$\tilde m=(mc_s)^3/(\pi^2 \hbar^3) \left[1-3/\big(32\pi \sqrt{m^2c_s^2 \beta}\big)\right]$, where the leading term, $(mc_s)^3/(\pi^2 \hbar^3)=8 n_c\sqrt{n_c a^3/\pi}$,
is the standard anomalous density for Bose gases \cite{Boudj,Griffin,Yuk}.
Remarkably, $\tilde m$ diverges as $\sim 1/\sqrt{ \beta} $ for small $\beta$ (see Fig.\ref{Fluc} dashed line).

Figure \ref{Fluc} shows that both functions $f$ and $h$ are decreasing with $\beta$ for any value of $(mc_s)^2$ indicating that QG effects 
lead to strongly reduce the condensed depletion and the anomalous density and thus, enhance the condensed fraction $n_c/n$. 
For instance, for $\beta =1/(mc_s)^2$ or equivalently $\beta_0 = (M_pc/mc_s)^2$, the depletion reduces to $\tilde n/n_c \approx 0.2 \sqrt{n_c a^3}$ whatever the value of $a$. 
The reason of such a decrease in $\tilde n$ and $\tilde m$ is most probably attributed to QG which acts as an extra force blocking interactions between bosons.
We observe also from the same figure that the anomalous density is larger than the noncondensed density in the whole range of $\beta$.

In weakly interacting Bose gases the fluctuations must be small. Therefore, the validity of the present theory at $T=0$ requires the condition:
\begin{equation}\label {Bog}
\sqrt{n_c a^3} f(\beta)  \ll 1,
\end{equation}
which differs by the factor $f(\beta) $ from the universal small parameter of the theory. 
The condition (\ref{Bog})  tells us that  the presence of QG conducts to the emergence of an ultradilute BEC since $f(\beta) $ is very small.

\subsection{Thermal fluctuations}

Let us now extend our results for the case of uniform BEC at finite temperature under the GUP. 
Evaluation of integrals (\ref{T:DH2}) and (\ref{T:DH3}) depends on the energy-momentum relation which unfortunately exhibits a divergent behavior at non-zero temperatures
for the usual Bogoliubov excitation energy. For $\beta\ll 1$, we expand the correction factor $(1+\beta P^2)^{-3}$ up to third-order as:
 $(1+\beta P^2)^{-3} = 1-3\beta P^2+ 6 (\beta P^2)^2-10 (\beta P^2)^3+ \cdots$ (see e.g.\cite{VaK, Zhang, Li}). 
This permits us to earn small  QG corrections to the condensate fluctuations and to its thermodynamics. 
However, when $\beta$ approaching unity, the system reaches the regime in which effects of QG become important. 
In such a case we should use numerical simulations in order to further analyze the influence of the GUP on the thermal properties of BEC.

At low temperatures $T \ll mc_s^2$, the main contribution to integrals (\ref{T:DH2}) and (\ref{T:DH3}) comes from the phonon region.
Using the identity $\int_0^{\infty}  x^j dx /(e^{x}-1) =\Gamma (j+1) \zeta (j+1)$, where $\Gamma (x)$ is the gamma function and $\zeta (x)$ is the Riemann zeta function, 
one obtains the temperature-dependence of the noncondensed and anomalous densities up to third-order in $\beta$:
\begin{align}\label {thfluc}
\tilde{n}_T =|\tilde{m}_T | = \frac{mT^2}{12\hbar^3 c_s} {\cal F} (\beta,T),
\end{align}
where
\begin{align}\label {thfluc}
{\cal F} (\beta ,T) &= 1-\frac{6\pi^2\beta}{5} \left(\frac{T}{c_s}\right)^2 + \frac{128\pi^4\beta^2}{21} \left(\frac{T}{c_s}\right)^4 \nonumber\\
&-16\pi^6 \beta^3 \left(\frac{T}{c_s}\right)^6+\cdots.\nonumber
\end{align}
Unlike the zero-temperature case, the function ${\cal F} (\beta ,T)$ depends on boson mass and varies as $m^{2\alpha} (na)^{-\alpha} $, where $\alpha>0$.
It is obvious that for $\beta=0$,  Eq.(\ref{thfluc}) coincides with the standard prediction for weakly-interacting BEC without considering QG which is $\propto T^2$.
This differs from the result of the ideal gas where $\tilde{n}_T$ behaves like $T^{3/2}$.
Equation (\ref{thfluc}) shows that at low temperatures $\tilde{n}_T$ and $\tilde{m}_T$ are of the same order of magnitude but with opposite signs. 

\begin{figure}
\centering 
\includegraphics[scale=0.45, angle=0] {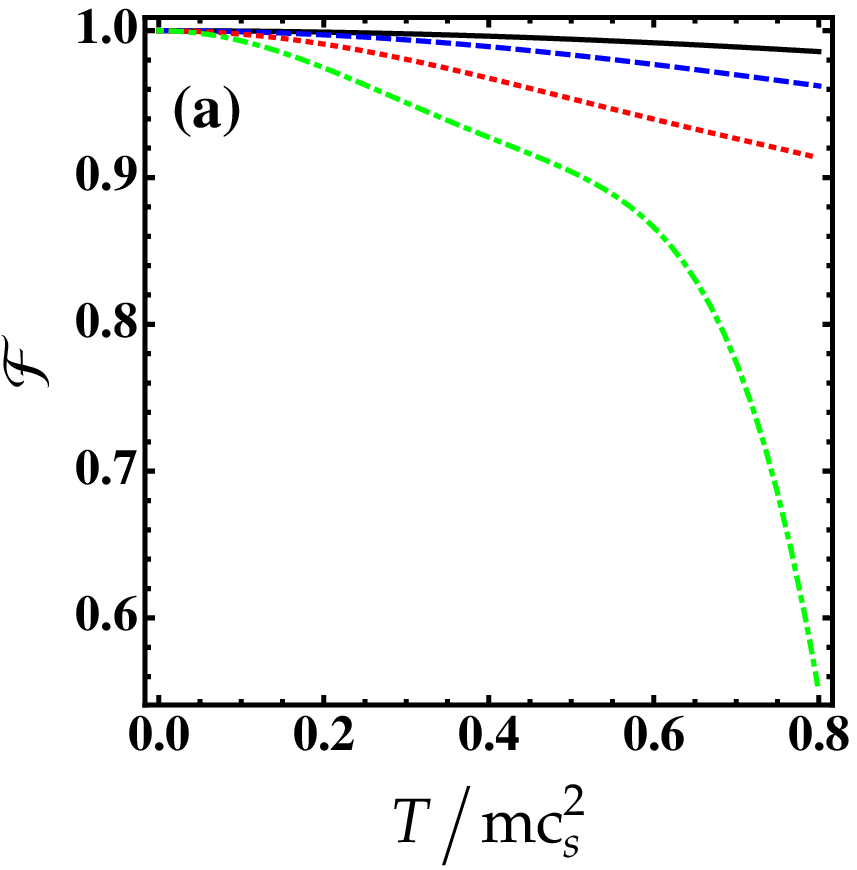}
\includegraphics[scale=0.47, angle=0] {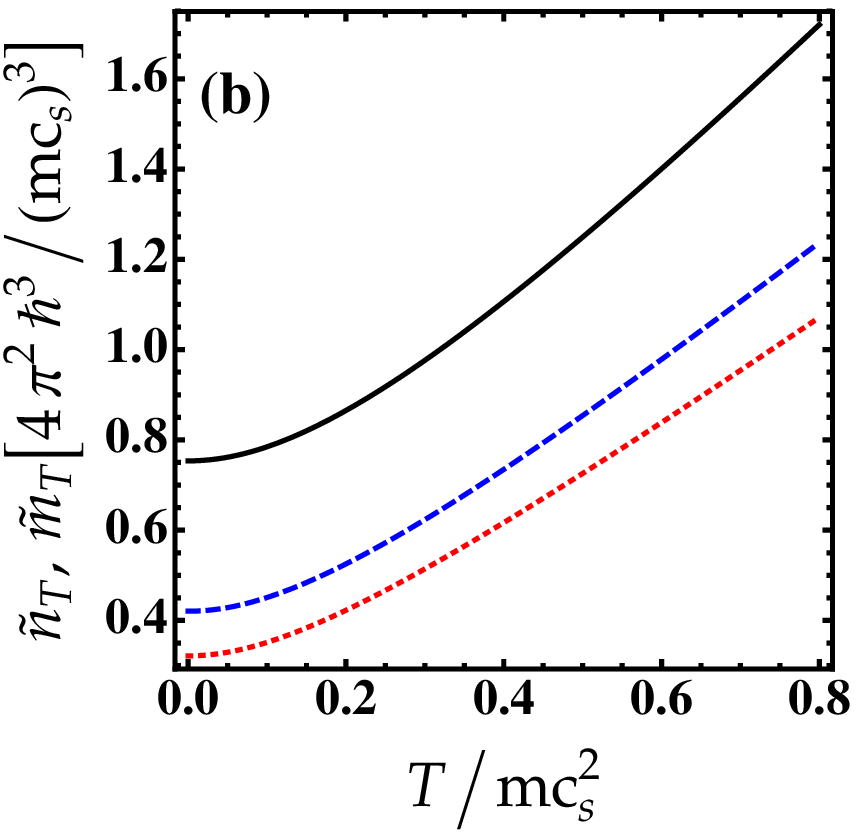}
\caption { (a) Deformation function ${\cal F}(\beta,T)$ which governs the dependence of thermal fluctuations 
as a function of the temperature $T/mc_s^2$ for several values of the deformation parameter  $\beta$ in units of $(mc_s)^2$.
Solid line: $\beta=0.002/(mc_s)^2 $. Dashed line: $\beta=0.006/(mc_s)^2 $. Dotted line: $\beta=0.02/(mc_s)^2 $. Dotted-dashed line:$\beta=0.06/(mc_s)^2$.
(b) Numerical simulation of thermal contribution from integrals (\ref{T:DH2}) and (\ref{T:DH3}).
Solid line: $\beta=0.5/(mc_s)^2 $. Dashed line: $\beta=0.8/(mc_s)^2$. Dotted line: $\beta=1/(mc_s)^2 $. }
\label{TFluc}
\end{figure}

Figure \ref{TFluc}.(a) depicts that as $\beta$ and $T$ rise, the function ${\cal F}(\beta,T)$ decreases which means that
the effects of QG are crucial specifically at low temperatures $T\ll T_c$. This behavior holds true also in ideal Bose gases \cite{Das}. 
For instance, in the case of  ${}^{133}$Cs BEC with parameters: $a=450 a_0$ ($a_0$ is the Bohr radius)\cite{Grim}, $n=10^{20}$ m$^{-3}$,
the function ${\cal F}(\beta,T)$ lowers by $\sim 13\%$ from $\beta =0.006/(mc_s)^2$ to $\beta =0.06/(mc_s)^2 $ at temperature $T \simeq 0.27 T_c^0$. 
Whereas, it decreases by $\sim 44\%$ at  $T \simeq 0.33 T_c^0$ for the same values of $\beta$.
This confirms that the condensate under the GUP still survives even in the limit of high temperature.
Here we use the fact that $T/mc_s^2 =(T/T_c^0)/[(na^3)^{1/3} 2\zeta^{2/3} (3/2)] $, where
$T_c^0 = 2\pi \hbar^2 [n/\zeta (3/2)]^{2/3} /m$ is the critical temperature of an ideal Bose gas. 
It is worth stressing that $T_c$ strongly depends on the density and on the interaction strength. 
 
The situation is quite different for large $\beta$ where the numerical simulation of integrals (\ref{T:DH2}) and (\ref{T:DH3}) clearly shows that
the thermal contribution to the noncondensed and anomalous densities is increasing with temperature and decreasing with $\beta$ regardless of the value of $(mc_s)^2$ (see Fig.\ref{TFluc}.(b)).
At temperatures $T> 0.25 T_c$, the thermal fluctuations of the condensate are important whatever the value of $\beta$. 

At temperatures $T\ll mc_s^2 $, the small parameter of the theory turns out to be given as: 
\begin{equation}\label {BogT}
{\cal F} (\beta ,T) (T/ mc_s^2) (na^3)^{1/2} \ll 1.
\end{equation}
The occurrence of the extra factor [${\cal F} (\beta ,T) (T/ mc_s^2) $] is due to the interplay of QG and the thermal fluctuations.
For ${\cal F} (\beta ,T)=1$, the criterion (\ref{BogT}) reduces to that obtained for BEC without considering the GUP \cite{Gora}.

At $T \gg mc_s^2$ where the main contribution to integrals (\ref{T:DH2}) and (\ref{T:DH3}) comes from the single particle excitations,
there is copious evidence that $\tilde{n}_T$ becomes identical to the noncondensed density of an ideal Bose gas. 
This implies that corrections to all thermodynamic quantities are closer to the values obtained for an ideal Bose gas.
However, the anomalous density being proportional to the condensed density, tend to zero together and hence,
their contributions become negligeably small \cite{Boudj,Yuk, Griffin}.

\section{Thermodynamic quantities} \label{Therm}

In this section, we analyze the influence of the GUP on the thermodynamic properties of weakly interacting Bose gases.

The chemical potential can be easily obtained from Eq.(\ref{chim}),
\begin{equation}\label {chim1}
\mu = gn +\frac{g (mc_s)^3}{4 \pi^2 \hbar^3} \left[f(\beta)-h(\beta)\right].
\end{equation}
Using the asymptotic behavior of the functions $f$ and $h$ for $\beta \rightarrow 0$, the 
equation of state (\ref{chim1}) becomes $\mu = gn + gn_c (32/3) \sqrt{na^3/\pi}[1-9\pi/(128 \sqrt{m^2c_s^2\beta})]$. 
The quantum corrections term $gn_c (32/3) \sqrt{na^3/\pi}$ was first derived by the LHY \cite{LHY}.  Again, the chemical potential decays as $\sim 1/\sqrt{\beta}$
for small $\beta$.

The ground-state energy of ultracold Bose gases in the presence of the GUP is defined as:
\begin{align}\label {energ}
\frac{E}{V}=&\frac{gn^2}{2}+\frac{1}{4 \pi^2 \hbar^3}\int \frac{P^2d  P} {(1+\beta P^2)^3} \bigg[\varepsilon_{P} - E_P-mc_s^2 \bigg].
\end{align} 
Unlike the ordinary BEC, the ground-state energy (\ref{energ}) is safe from the ultraviolet divergence due to the presence of a minimal length.
After a straightforward calculation, we find
\begin{equation}\label {energ1}
\frac{E}{V}=\frac{1}{2} g n^2 + \frac{m^4 c_s^5}{4 \pi^2\hbar^3} S(\beta), 
\end{equation} 
where  the energy deformation function $S(\beta)$ is given by
\begin{widetext}
\begin{align}\label{DF3} 
S(\beta)&=-\frac{1}{32 (1-4 m^2c_s^2\beta )^{3/2} m^2c_s^2\beta ^{5/2}} \bigg\{ \sqrt{1-4 m^2c_s^2\beta } \bigg[4 \sqrt{m^2c_s^2\beta } (8 m^2c_s^2\beta -3) 
+\pi  \big(3-2 m^2c_s^2\beta  (4 m^2c_s^2\beta +5)\big)\bigg]\\
&+(32 m^2c_s^2\beta -6) \arccos\left(2 \sqrt{m^2c_s^2\beta }\right) \bigg\}. \nonumber
\end{align} 
\end{widetext}
For $\beta \rightarrow 0$, $S(\beta)\approx -3\pi/(16 \sqrt{m^2c_s^2\beta}) +32/15$ 
hence, Eq.(\ref{energ1}) simplifies to $E/V=g n^2/2 \big[1+ (128 gn^2/15) \sqrt{na^3/\pi} \big(1-45\pi/512 \sqrt{m^2c_s^2\beta}\big)\big]$.
We see from Fig.\ref{Fluc} (dotted line) that the function $S(\beta)$ is increasing with $\beta$ and remains negative
signaling that the effects of QG  tend to lower the ground-state energy.
This emphasizes that the subleading term in Eq.(\ref{energ1}) which arises from the LHY quantum fluctuations
and QG corrections is much smaller than the leading contribution to the energy which originates from interactions.

In the frame of our formalism, the free energy can be written as:
\begin{equation}\label {fergy}
F=E+\frac{T}{2 \pi^2 \hbar^3}\int \frac{P^2 d P} {(1+\beta P^2)^3}   \ln\left(\frac{2} {\sqrt{I_P}+1}\right),
\end{equation}
where $E$ is the ground-state energy given in Eq.(\ref{energ1}).
Integrating the subleading term in Eq.(\ref{fergy}) for temperatures satisfying the inequality $T \ll mc_s^2$, we obtain up to third-order in $\beta$:
\begin{align}\label {fergy1}
F=E+\frac{\pi^2 T^4}{90\hbar^3 c_s^3} {\cal H} (\beta, T),
\end{align}
where
\begin{align}
{\cal H} (\beta, T)&= -1+\frac{24\pi^4\beta}{7} \left(\frac{T}{c_s}\right)^2- \frac{144\pi^6\beta^2}{7} \left(\frac{T}{c_s}\right)^4 \nonumber\\
&+\frac{6400\pi^8 \beta^3}{33} \left(\frac{T}{c_s}\right)^6-\cdots. \nonumber
\end{align}
Expression (\ref{fergy1}) clearly shows that for $\beta=0$, one recovers the famous $T^4$-law for the free energy, in contrast with the $T^{5/2}$
behavior found for the non-interacting Bose gas \cite{peth}.
 
\begin{figure}
\centering 
\includegraphics[scale=0.45, angle=0] {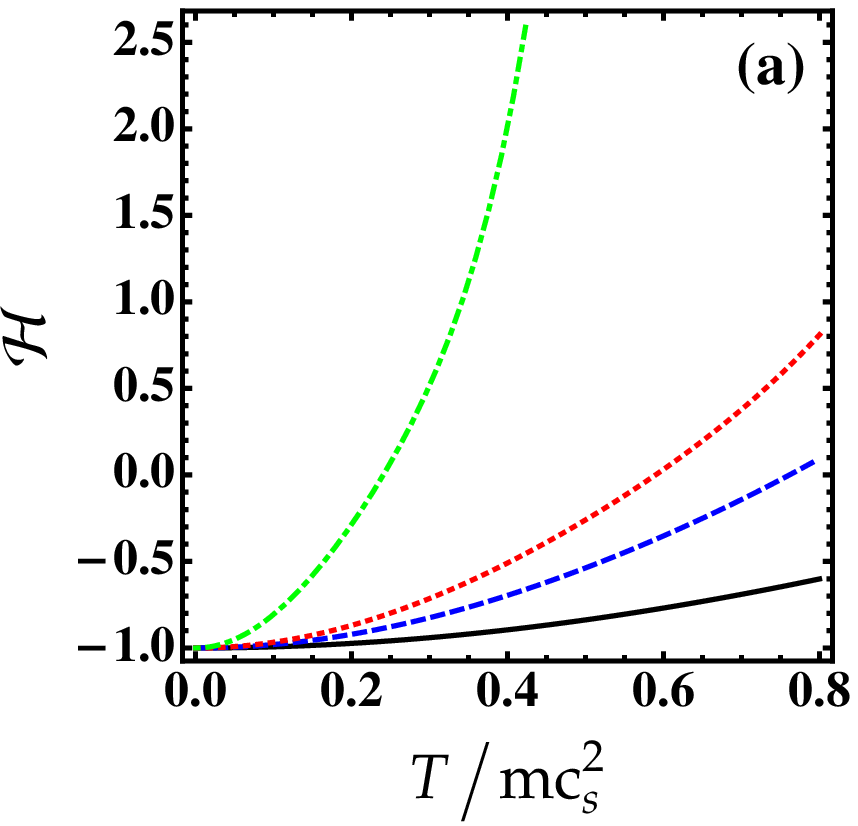}
\includegraphics[scale=0.48, angle=0] {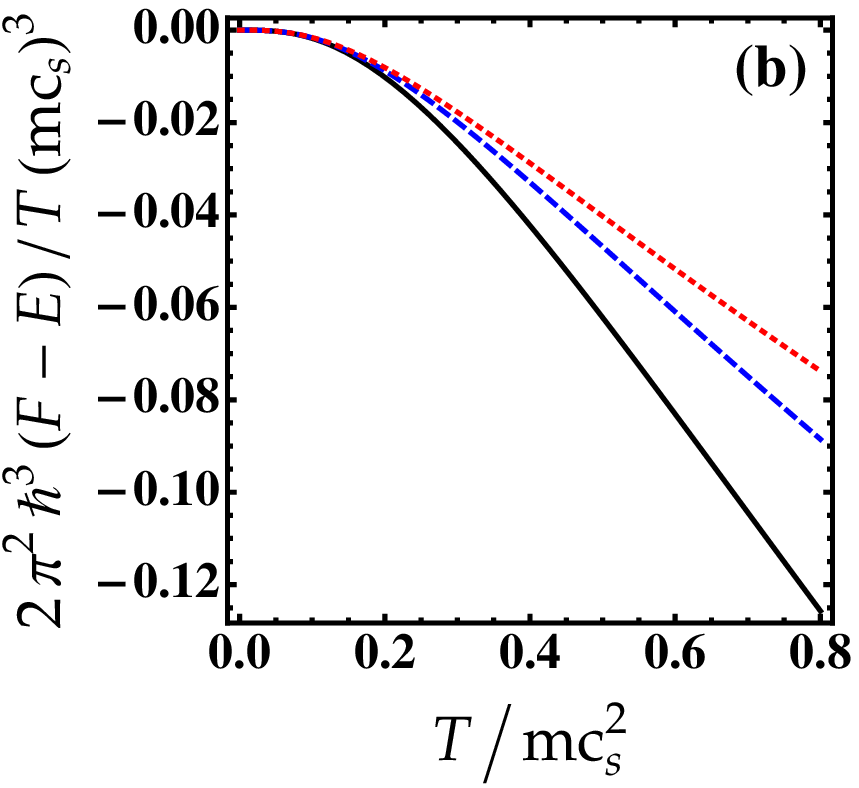}
\caption { (a) Deformation function ${\cal H}(\beta,T)$ which governs the dependence of thermal contribution to the free energy
as a function of the temperature for several values of the deformation parameter $ \beta$ in units of $(mc_s)^2$.
Solid line: $\beta=0.002/(mc_s)^2$. Dashed line: $\beta=0.006/(mc_s)^2$. Dotted line: $\beta=0.02/(mc_s)^2$. Dotted-dashed line: $\beta=0.06/(mc_s)^2$.
(b) Numerical simulation of thermal contribution of the free energy from Eq.(\ref{fergy}).
Solid line: $\beta=0.5/(mc_s)^2$. Dashed line: $\beta=0.8/(mc_s)^2$. Dotted line: $\beta=1/(mc_s)^2$.}
\label{Fregy}
\end{figure}

The function ${\cal H} (\beta, T)$ is displayed in Fig.\ref{Fregy} (a).
We see that for small $\beta$, it increases monotonically with $T/mc_s^2$ and remains negative.
As $\beta$ increases ${\cal H} (\beta, T)$ changes its character from positive at low temperatures to negative at relatively higher $T$.
In this case, the system is viewed as being dominated by QG effects reflecting the formation of an unstable BEC.

In Fig.\ref{Fregy}.(b) we represent the numerical simulation of the free energy for large $\beta$  in units of $(mc_s)^2$. 
We observe that corrections due to QG remain negative and tiny except at very low temperatures $T \lesssim 0.08 T_c$.
Here the negative term is smaller than the positive one signaling that the system is in its stable state.

\section{Superfluidity} \label{SupGUP}

Fluctuations due to QG may lead also to modify the superfluid density. 
In our formalism it is defined as \cite{Boudj3}
\begin{equation}\label {supf}
n_s=1-n_n=1-\frac{2Q}{3T},
\end{equation}
where  $n_n=2Q/3T$ accounts for the normal density of the Bose-condensed liquid, and
$Q$ is the dissipated heat defined in an equilibrium system through the average of the total kinetic energy per particle. 
This yields
\begin{equation}\label {supf1}
Q= \frac{1}{2 \pi^2 \hbar^3}\int \frac{P^2 d P} {(1+\beta P^2)^3}  \frac{ E_P (I_P-1)}{4}. 
\end{equation}
In the absence of QG the dissipated heat (\ref{supf1}) becomes identical to that of Ref.\cite{Yuk}.
Such two-fluid hydrodynamics developed by Landau and Khalatnikov \cite{Khal,LL9} incorporates the motion 
of both the superfluid background (gravitational field) and excitations (matter). 
This is equivalent of the Einstein equations which encompass both gravity and matter \cite{Volov}.

Again at $T \ll mc_s^2$, a straightforward calculation up to third-order in $\beta$ gives for the superfluid density:
\begin{equation}\label {supf2}
n_s=1-\frac{2\pi^2T^4}{45 m\hbar^3 c_s^5} {\cal S}(\beta,T),
\end{equation}
where 
\begin{align}
{\cal S}(\beta,T)&= 1-\frac{60 \pi^4 \beta}{7}\left(\frac{T}{c_s}\right)^2+96 \pi^6 \beta ^2\left(\frac{T}{c_s}\right)^4\nonumber\\
&-\frac{128000\pi^8 \beta^3}{11} \left(\frac{T}{c_s}\right)^6+\cdots. \nonumber
\end{align}
For $\beta =0$, the superfluid density returns to the seminal Landau's formula $n_s=1-2\pi^2T^4/(45 m\hbar^3 c_s^5)$ \cite{LL9}.

\begin{figure}
\centering 
\includegraphics[scale=0.45, angle=0] {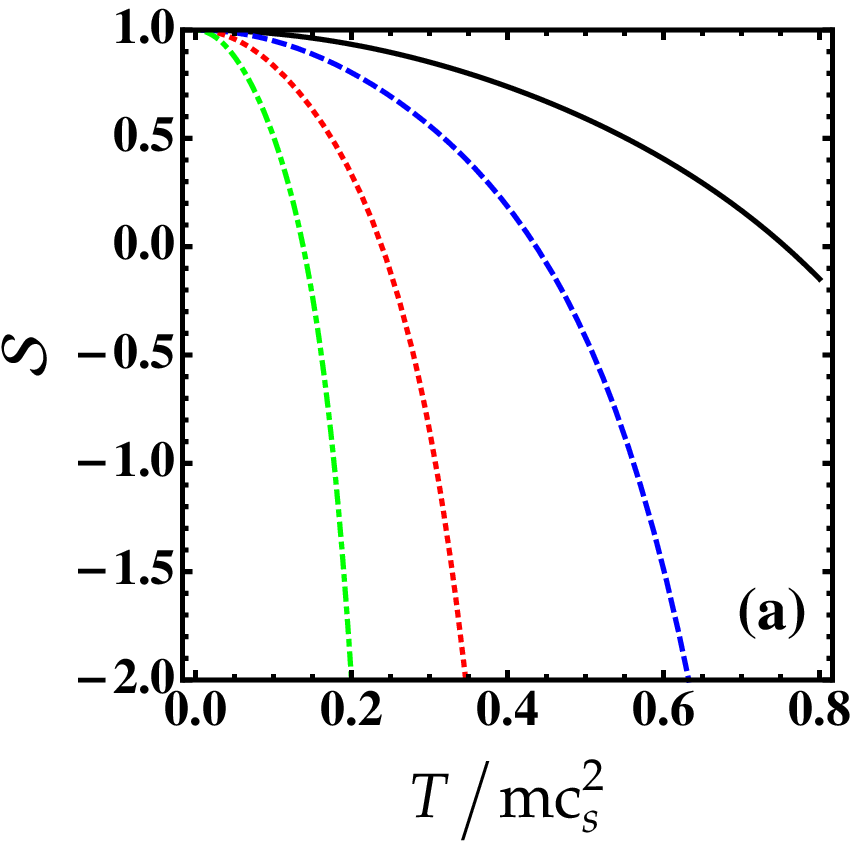}
\includegraphics[scale=0.46, angle=0] {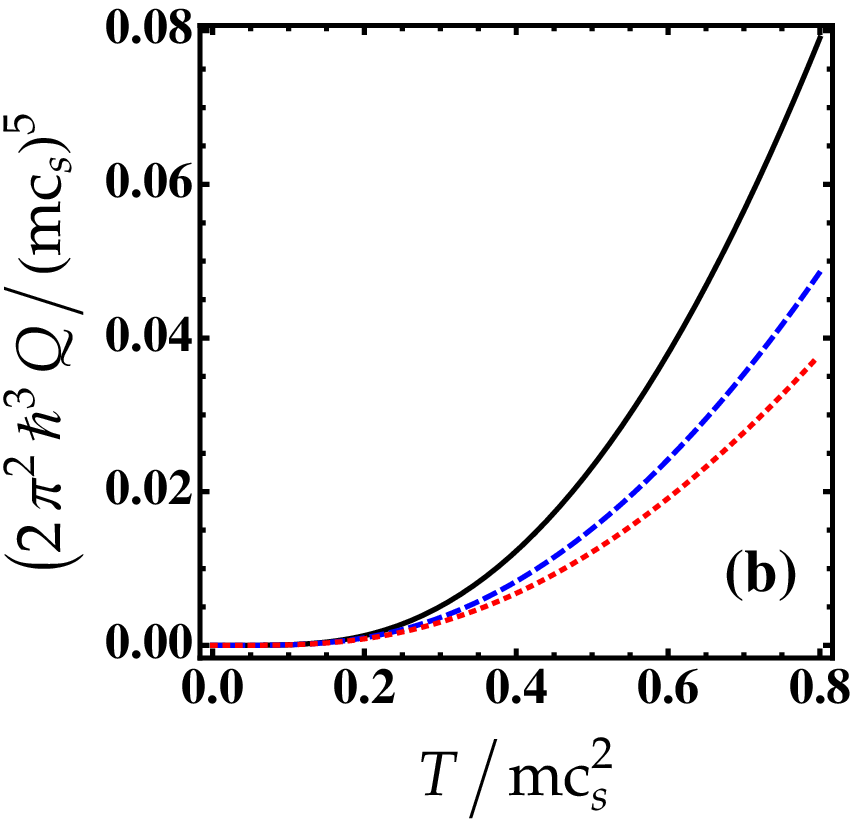}
\caption { (a) Deformation function ${\cal S}(\beta,T)$ which governs the dependence of normal component of the superfluid density
as a function of the temperature for several values of the deformation parameter $\beta$ in units of $(mc_s)^2$.
Solid line: $\beta=0.002/(mc_s)^2$. Dashed line: $\beta=0.006/(mc_s)^2$. Dotted line: $\beta=0.02/(mc_s)^2$. Dotted-dashed line: $\beta=0.06/(mc_s)^2$.
(b) Numerical simulation of the normal density of the superfluid from Eq.(\ref{supf1}).
Solid line: $\beta=0.5/(mc_s)^2$. Dashed line: $\beta=0.8/(mc_s)^2$. Dotted line: $\beta=1/(mc_s)^2$.}
\label{Supf}
\end{figure}
Figure \ref{Supf}.(a) shows that for fixed $(mc_s)^2$, the function ${\cal S}$ is decreasing with increasing $\beta$ which may lead to strongly reduce the normal density of the superfluid. 
Therefore, QG may enhance  the superfluid density even for small $\beta$. Another important remark is that at $T=0$, the whole liquid is superfluid i.e. $n_s=n$. 

Corrections to the normal density due to QG for large $\beta$ are computed numerically.
We can see from Fig.\ref{Supf}.(b) that as $\beta$ gets larger (approaches to unity), $n_n$ diminishes. 
The normal density remains minor even at higher temperatures in contrast with the thermal cloud of the condensate.

Notice that at $T \gg mc_s^2$, the normal density agrees with the noncondensed density of an ideal Bose gas at least for small $\beta$.

\section{Experimental Test of Quantum Gravity} \label{Exp}

In this section we discuss the possible experimental tests of our theoretical predictions that include QG effects.
The mesurement of quantum depletion of an interacting homogeneous BEC has been recently reported in \cite{Lopes}.
In such an experiment, the quantum depletion is of the order of 1\%, while by largely increasing the scattering
length using Feshbach resonances, $\tilde n$ as increases as about 10\% \cite{Lopes}.

\begin{table}[h!]
\begin{center}
\begin{tabular} { cccc cc} 
 \hline \hline\\
                        & $n$  & $a$   & $\tilde n/n $  & $\beta_0$ & $\Delta X_{\text{min}}$ \\ 
 \hline\\
 ${}^{39}$K      & 3.5$\times 10^{17}$m$^{-3}$ & 200 $a_0$  & 0.001   & 4.2$\times 10^{57}$   &  $\sim 1.8 \mu$m \\   \\
                       & 3.5$\times 10^{17}$m$^{-3}$ & 3000 $a_0$ & 0.1      & 7.5$\times 10^{56}$   &  $\sim 0.7 \mu$m \\                 
 \hline \\
 ${}^{133}$Cs  & 1.2$\times 10^{18}$m$^{-3}$ & 450 $a_0$  & 0.01     &  $1.3 \times 10^{57}$  &  $\sim 1 \mu$m \\ \\
                       & 1.2$\times 10^{18}$m$^{-3}$ & 2000 $a_0$  & 0.1    & $ 3.3 \times 10^{56}$   &  $\sim0.5 \mu$m\\
 \hline\hline
\end{tabular}
\end{center}
\caption{Typical values of $\beta_0$ and $\Delta X_{\text{min}}$ obtained from Eq.(\ref{GUP0}) for ${}^{39}$K BEC \cite{Lopes} and ${}^{133}$Cs BEC \cite{Grim}. 
Here the values of $a$ can be adjusted via Feshbach resonances \cite{Lopes, Grim1}.}
\label{table:1}
\end{table}

Table \ref{table:1} shows typical values of $\beta_0$ and $\Delta X_{\text{min}}$ obtained from Eq.(\ref{GUP0}) for ${}^{39}$K BEC \cite{Lopes} and ${}^{133}$Cs BEC \cite{Grim}.
We see that for a sufficiently dilute Bose gas, our model predicts much more large QG parameter $\beta_0\sim 10^{56}-10^{57}$  and 
a minimum measurable length $\Delta X_{\text{min}}$ around $ 1\mu$m.
For large depletion, our results for the QG parameter ($\beta_0\sim 10^{56}$)  are comparable to those obtained recently for optical systems \cite{Bra}.
Whereas, our findings for a minimum length are higher by 100 of order of magnitude than those predicted for graphene $\Delta X_{\text{min}} \sim 2.3$ nm \cite{Menc}.

The detection threshold for quantum depletion is of the order $10^{-3}$ for weak interactions and about $\sim 10^{-2}$ for relatively strong interactions \cite{Lopes}.
Therefore, if one is to deal with actual laboratory measurements of $\tilde n$ including QG effects, the deformation parameter must be 
bounded as $\beta_0 < 10^{57}$. 
It is important to mention here that the depletion corrections can be modified merely by changing the $s$-wave scattering length $a$, and hence improve the bound on the QG parameter.

\section{Conclusions}\label{conc}

We studied QG effects due to the GUP on the properties of weakly interacting homogeneous Bose gases at both zero and finite temperatures
using  the self-consistent TDHFB theory. The developed approach which is a combined theory of the HFB formalism and the GUP may be useful to
test QG in real ultracold dilute Bose gases. 
We showed that corrections due to the presence of a minimal length provide an extra term modifying the condensate depletion, the anomalous density, the chemical potential, 
the ground-state energy, the free energy, and the normal density of the superfluid. 
Our results pointed out also that the condensate depletion is tunable by changing the QG parameter. 
When the GUP is not considered ($\beta=0$), the obtained results become consistent with those existing in the literature.
The competition between quantum fluctuations induced by interactions and  QG effects may also shift the critical temperature allowing the
formation of BEC and Bose superfluids even at higher temperatures.
Finally, we showed that the experimental observation of such a system requires much more pronounced QG parameter.
We expect that our findings will provide deeper insights into the universal properties of the dilute Bose gas under the GUP.

We believe that this work opens new prospects to the unexplored frontier of microscopic theory which will supply a real opportunity for a future tabletop test for 
unification of quantum theory and general relativity with a BEC.
In our future work, we aim to investigate QG effects in dipolar BEC using our HFB model.
Ultracold atoms with dipole-dipole interactions \cite{Boudj15} which have the same form as the quantum gravitational interactions would be interesting  
to distinguish the QG signal from electromagnetic force \cite{Howl}, and hence enable studies for testing QG. 
\\

\subsection*{Data availability statement}
The data generated and/or analyzed during the current study are not publicly available for legal/ethical reasons
but are available from the corresponding author on reasonable request.

\end{document}